\pdfoutput=1

\documentclass{appolb}

\usepackage{amssymb}
\usepackage{hyperref}
\usepackage{graphicx}
\usepackage{amsmath}
\usepackage[numbers,sectionbib,sort&compress]{natbib}

\newcommand{\br}[1]{\langle #1\rangle}
\newcommand{\bn}[1]{\left( #1\right)}

\newcommand{\var}[1]{{\rm var}(#1)}
\newcommand{\cov}[2]{{\rm cov}(#1,#2)}

\begin{document}

\title{Forward-backward multiplicity correlations at the LHC from independent sources%
\thanks{Supported by the Polish National Science Center grant 2015/19/B/ST2/00937}}

\author{Adam Olszewski$^{1}$\thanks{adam.olszewski.fiz@gmail.com} and Wojciech Broniowski$^{1,2}$\thanks{Wojciech.Broniowski@ifj.edu.pl}
\address{$^{1}$Institute of Physics, Jan Kochanowski University, PL-25406~Kielce, Poland}
\address{$^{2}$The H. Niewodnicza\'nski Institute of Nuclear Physics, Polish Academy of Sciences, PL-31342~Cracow, Poland}}

\maketitle

\begin{abstract}
It is argued that the superposition approach, where partons are independently emitted from longitudinally extended sources in the early stage, is fully compatible with the experimental results for the forward-backward multiplicity correlations in Pb+Pb collisions at $\sqrt{s_{NN}}=2.76 \textrm{~TeV}$. The pertinent correlation analysis is based on the PhD Thesis of Ref.~\cite{Sputowska:2016xxx}, which includes an unpublished analysis of data taken by the ALICE Collaboration. Our calculations show that in the experimentally covered pseudorapidity range $\Delta \eta=1.2$, the initial sources in the backward and forward bins are maximally correlated, which complies to the string-like interpretation of the underlying  early-stage production mechanism.
\end{abstract}

PACS: 5.75.-q, 25.75Gz, 25.75.Ld

\bigskip
  
\section{Introduction \label{sec:intro}}

In this paper we use the method developed in Ref.~\cite{Olszewski:2013qwa,Olszewski:2015xba} 
to confirm that the mechanism of early particle production at the Large Hadron Collider (LHC) may be understood, to a good
approximation, in terms of emission from independent sources which extend over a wide longitudinal range. Our 
analysis is performed with the help of simple formulas from Ref.~\cite{Olszewski:2013qwa} for the correlation coefficients. 
It uses the data taken by the ALICE Collaboration for Pb+Pb collisions at $\sqrt{s_{NN}}=2.76 \textrm{~TeV}$ in the form presented 
in the PhD thesis by I.~Sputowska~\cite{Sputowska:2016xxx}. 

As is well known, the long-range rapidity correlations in hadronic collision experiments 
reveal information on the dynamics and evolution of the system 
in its earliest partonic phase. Experimentally, the multiplicity correlations 
in early $pp$ and $p\bar{p}$ collisions~\cite{Uhlig:1977dc,Alpgard:1983xp,Alner:1987wb,Ansorge:1988fg,Derado:1988ba,Alexopoulos:1995ft} 
and nuclear collisions~\cite{Bachler:1992psr,Akiba:1997yg} were
followed by the relativistic heavy-ion and $pp$ experiments at RHIC~\cite{Back:2006id,Abelev:2009ag,Tarnowsky:2010qp} and the 
LHC~\cite{ATLAS:2012as,Jia:2015jga,Adam:2015mya,Aaboud:2016jnr,Sputowska:2016xxx}. Physical pictures, models, and theoretical methods have been constructed along the quest to 
understand the data~\cite{Capella:1978rg,Kaidalov:1982xe,Chou:1984wp,Capella:1992yb,%
Amelin:1994mf,Braun:2000cc,Giovannini:2002za,Braun:2003fn,Brogueira:2007ub,%
Armesto:2007ia,Armesto:2006bv,Vechernin:2007zza,Braun:2007rf,%
Konchakovski:2008cf,Bzdak:2009dr,Lappi:2009vb,Bozek:2010vz,deDeus:2010id,Bialas:2011xk,Bialas:2011vj,%
Bzdak:2011nb,Bzdak:2012tp,Vechernin:2012bz,Bialas:2013xea, De:2013bta,Ma:2014pva,Bzdak:2015dja,Bzdak:2015eii,Vechernin:2015upa}. 

The basic assumptions of the applied {\em superposition framework} are following~\cite{Olszewski:2013qwa,Olszewski:2015xba}:
\begin{enumerate}
 \item[(a)] Particle emission occurs independently from longitudinally extended sources.
 \item[(b)] The forward (F) and backward (B) bins are sufficiently well separated in pseudorapidity, such that the transition from the 
 initial state to the final hadron distribution does not cause mixing between particles belonging to the F and B bins.
\end{enumerate}
Actually, our approach takes into account three stages typically distinguished in the evolution of the system: 
1)~early production of initial particles (forming an entropy density) from sources, 2)~hydrodynamic or transport evolution in 
the intermediate phase, and finally 3)~production of hadrons and their subsequent registration in detectors.

Our derivation assumes for simplicity a single type of sources. In Appendix~\ref{app:multi} we show how 
and under what conditions the 
model may be generalized to a case with multiple types of sources

\section{Formulas} \label{sec:formulas}

As explained in detail in Refs.~\cite{Olszewski:2013qwa,Olszewski:2015xba} (cf. also Appendix~\ref{app:supp} in the present work), 
the above-mentioned stages 1) and 3) involve, from the statistical point 
of view, folding of statistical distributions, whereas stage 2) results in a linear transformation of the particle (fluid) density. 
The three stages may be combined to yield a very simple ``pocket'' formula involving only one free parameter, relating the correlation of the 
initial sources $s_F$ and $s_B$ in the F and B bins in spatial rapidity, denoted as $\rho(s_F,s_B)$, to statistical quantities accessible experimentally. 
These quantities are
the correlation of the numbers of charged hadrons $n_F$ and $n_B$ in the experimental F and B bins in pseudorapidity, denoted as 
$\rho(n_F,n_B)$ (a.k.a. the $b$ coefficient), and the scaled variances of multiplicities in the F and B bins, $\omega(n_F)$ and $\omega(n_B)$.
For symmetric collisions and for symmetrically arranged pseudorapidity bins $\omega(n_F)=\omega(n_B)\equiv \omega(n_A)$, and we have 
(see Appendix~\ref{app:supp})
\begin{eqnarray}
\rho(s_F,s_B) &=& \frac{\rho(n_F,n_B)}{1-\frac{\delta}{\omega\bn{n_A}}}, \label{eq:rhos}
\end{eqnarray}
where $\rho$ stands for Pearson's correlation coefficient, $\omega$ denotes the scaled variance, and 
$\delta$ is a phenomenological constant, whose anatomy is discussed in Appendix~\ref{app:supp}. An important feature is 
that $\delta$ does not depend on the rapidity separation of the F and B bins, nor (to a good approximation) on the centrality of the collision. Thus, for a given 
experimental setup (energy of the collision, width of the bins in pseudorapidity, detector acceptance) it is {\em constant}. 
We can rearrange Eq.~(\ref{eq:rhos}) to extract $\delta$:
\begin{eqnarray}
\delta &=& \omega\bn{n_A}\bn{1-\frac{\rho(n_F,n_B)}{\rho(s_F,s_B)}}. \label{eq:omr1}
\end{eqnarray}

It should be stressed that relations~(\ref{eq:rhos}-\ref{eq:omr1}) originate solely from assumptions (a) and (b) specified above and hold 
for any experimental data sample (e.g., any centrality cut). Thus their 
verification directly checks assumptions (a) and (b). Two straightforward tests emerge here, each based on one of the above formulas. 
First, we may use Eq.~(\ref{eq:omr1}) with the experimental data for $\rho(n_F,n_B)$ and  $\omega\bn{n_A}$, as well 
as with the assumption $\rho(s_F,s_B)=1$ which should hold for not too large bin separations $\Delta \eta$. If thus obtained
$\delta$ is indeed constant, the test is passed and the superposition model works. 
Second, we may use a suitably chosen constant value of $\delta$ in Eq.~(\ref{eq:rhos})
and obtain $\rho(s_F,s_B)$ at various centralities and bin separations $\Delta \eta$.  

\section{Results \label{sec:results}}

\begin{figure}[tb]
\begin{center}
\includegraphics[scale=.6]{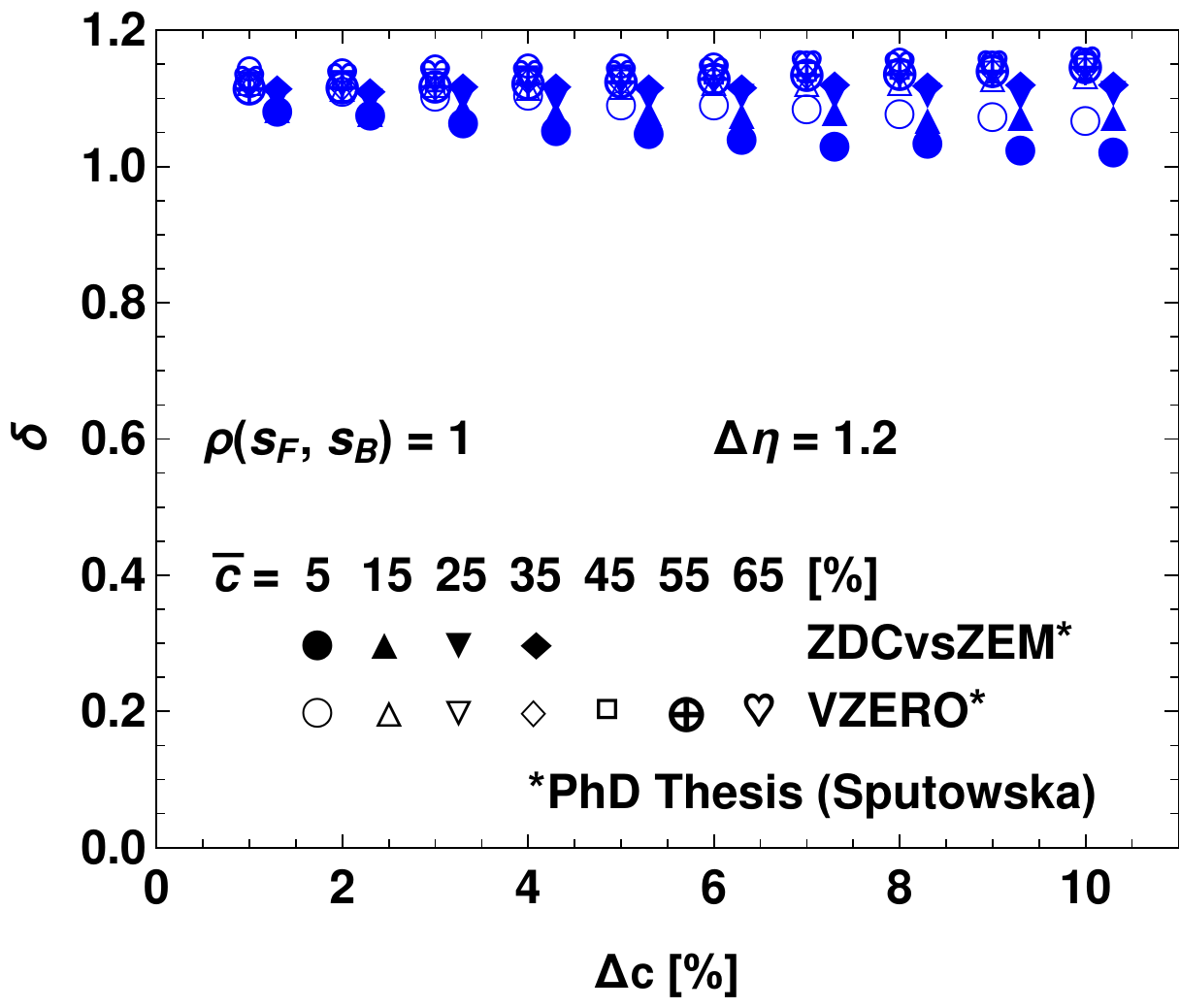}
\caption{Values of the $\delta$ parameter obtained from Eq.~(\ref{eq:omr1}) with the data for 
Pb+Pb collisions at $\sqrt{s_{NN}}=2.76 \textrm{ TeV}$ recorded by the ALICE experiment, digitized by the authors from Figs.~(3.3,3.4) of the PhD thesis~\cite{Sputowska:2016xxx}. 
The result is plotted as a function of the width of the centrality bin, $\Delta c$, for several centralities of the center of the bin, $\bar c$, and for two centrality selection 
methods  of Ref.~\cite{Sputowska:2016xxx}: \texttt{VZERO} (empty symbols) and \texttt{ZDCvsZEM} (filled symbols). The very 
similar values of $\delta$ conform to the assumption of emission from independent 
longitudinally-extended sources which are maximally correlated over the pseudorapidity separation $\Delta \eta=1.2$ between 
the forward and backward bins, i.e., $\rho(s_F,s_B)=1$.
\label{fig:deltacen}}
\end{center}
\end{figure}

\begin{figure}[tb]
\begin{center}
\includegraphics[scale=.6]{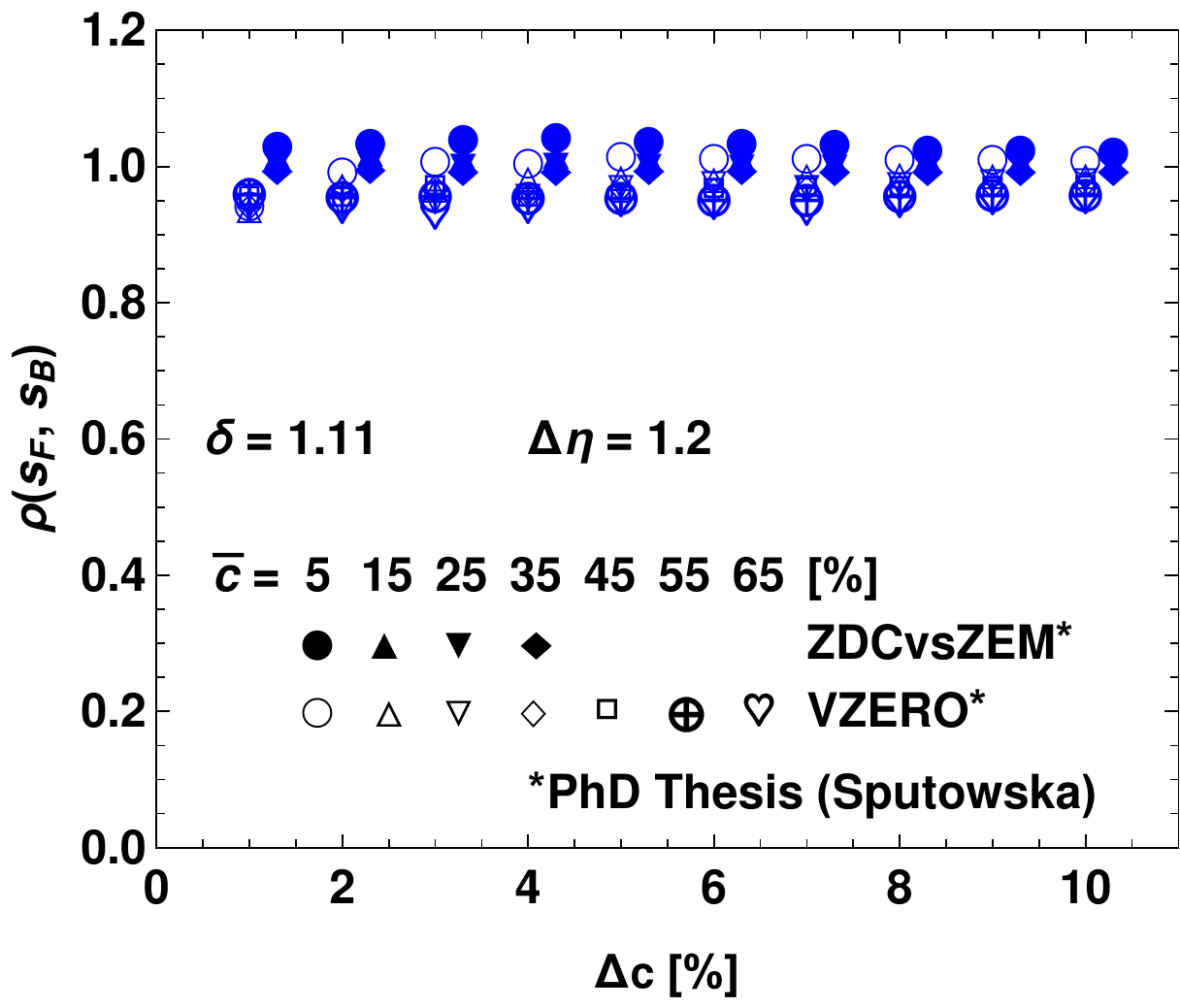}
\caption{Same as in Fig.~\ref{fig:deltacen} but for the forward-backward correlation of the sources $\rho(s_F,s_B)$ from Eq.~(\ref{eq:rhos}), 
evaluated with the average value of the 
superposition parameter $\delta=1.11$. 
\label{fig:fbsourcescen}}
\end{center}
\end{figure}

We begin presenting our results with the $\delta$ parameter obtained from Eq.~(\ref{eq:omr1}). The experimental 
quantities $\omega\bn{n_A}$ and $\rho\bn{n_F,n_B}$ are extracted from a manual digitalization of the points in Figs.~(3.3,3.4) published in 
the PhD Thesis by I.~Sputowska~\cite{Sputowska:2016xxx}. The ALICE measurements are carried out with two different methods 
of determining the centrality of the collision, \texttt{VZERO} (empty symbols) and \texttt{ZDCvsZEM} (filled symbols).
Essentially, the first method uses the multiplicity of hadrons in the central bin, whereas the other effectively determines the 
number of spectators (or participants) in the collision. 
We denote the center and the width of a centrality bin with $\bar c$ and $\Delta c$, respectively. 

Our values for $\delta$  are presented in Fig.~\ref{fig:deltacen} as a function of $\Delta c$ for the F and B bin separation 
$\Delta \eta=1.2$ (largest  accessible experimentally). This separation is sufficiently large to minimize the mixing between the bins during the evolution 
of the system (our assumption~(b)). At the same time, it is small enough to expect that the sources are maximally correlated, i.e., $\rho\bn{s_F,s_B}=1$. 
We note that the values for $\delta$ are within the band $1.1 \pm 0.1$ for both methods of the centrality determination and for various $\Delta c$ 
and $\bar c$. Taking into account the fact that $\omega\bn{n_A}$ and $\rho\bn{n_F,n_B}$ vary significantly (even up to factors of 5, cf. 
Figs.~(3.3,3.4) in Refs.~\cite{Sputowska:2016xxx}), the fact that the 
values of $\delta$ are almost constant is far from trivial and conforms to the superposition mechanism from independent sources.  

Of course, there are departures in $\delta$ from a strict constant value, and there is a number of factors which cause the effect:
some remnant mixing of the bins (caused, e.g., by partons emitted into distant pseudorapidities in the early stage, 
or resonance decays in the late stage), non-linearity of the hydrodynamic or transport evolution, leading to corrections
to the simple Eq.~(\ref{eq:hyd}). Also, there may be nonlinear effects in the early production mechanism, as present, e.g.,
in the mixed model~\cite{Kharzeev:2000ph}, where wounded nucleons~\cite{Bialas:1976ed} are amended with an admixture of binary collisions.
The fact that $\delta$ is to a good approximation constant shows that these effects are not very significant. 
We also note that the obtained values of $\delta$ are larger than $1$, which complies to the constraint~(\ref{eq:cons}). 

Next, in Fig.~\ref{fig:fbsourcescen} we present the result for the forward-backward correlation of the number of the initial sources, $\rho(s_F,s_B)$, 
obtained from the Eq.~(\ref{eq:rhos}), where we use the average value of $\delta$ from Fig.~\ref{fig:deltacen}, namely $\delta=1.1$. 
The correlation is plotted  as a function $\Delta c$ for various $\bar c$ and for the data with both \texttt{VZERO} and \texttt{ZDCvsZEM}
centrality determination methods for $\Delta \eta=1.2$, the same as used in Fig.~\ref{fig:deltacen}. 
We note that the resulting values for $\rho(s_F,s_B)$ are close to $1$, in accordance to the hypothesis of a maximum correlation 
of sources over a moderate pseudorapidity range. The fact that for certain cases  the points go slightly above $1$ (which is mathematically precluded
for the correlation coefficient) is caused by the above-listed effects modifying the simplest superposition model, as well as by 
experimental errors, not incorporated in our analysis. 

\begin{figure}
\begin{center}
\includegraphics[scale=.6]{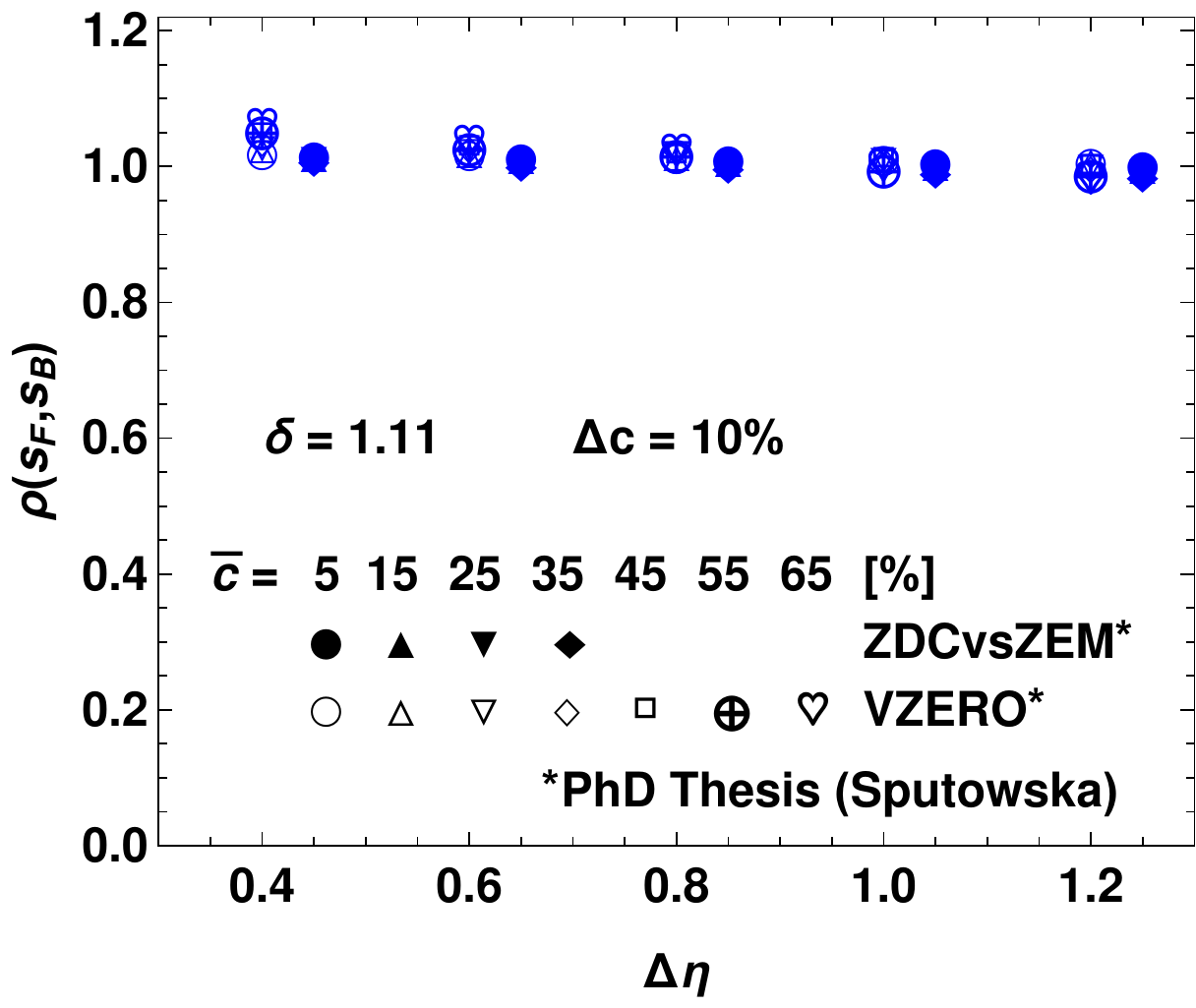}
\caption{Same as in Fig.~\ref{fig:fbsourcescen} but plotted 
for several values of centrality $\bar c$ as a function of the forward-backward pseudorapidity separation $\Delta \eta$.}
\label{fig:corretasfsb}
\end{center}
\end{figure}

Finally, in Fig.~\ref{fig:corretasfsb} we plot $\rho(s_F,s_B)$ as a function of the pseudorapidity separation $\Delta \eta$
for the case $\Delta c=10\%$. 
For this purpose the necessary data were digitized from Figs.~(3.1,3.6) of Ref.~\cite{Sputowska:2016xxx}. 
As before, we use $\delta=1.11$.
We note that in the covered range of $\Delta \eta$ the resulting $\rho(s_F,s_B)$ is very close to $1$ and independent of the 
centrality $\bar c$.

\section{Conclusions \label{sec:cls}}

The main result of our analysis is that the hadron production mechanism based on production from independent sources,
strongly correlated over the accessible pseudorapidity range, works very well in Pb+Pb collisions at the LHC. The key test here is 
the constant value of the $\delta$ parameter, as exhibited in Fig.~\ref{fig:deltacen}. It also shows 
that the data analysis based on standard measures of F-B correlations is by all means useful and allows us for 
access to physics questions of the particle production mechanism in ultra-relativistic heavy-ion collisions. 
Note that the usefulness for the tests of the superposition mechanism explored here 
holds despite the effect of {\em centrality fluctuations}, which may be reduced through the use 
of other more elaborate correlation measures~\cite{Bzdak:2012tp,De:2013bta,Bhalerao:2014mua,Bzdak:2015dja,Jia:2015jga,He:2016qjs,He:2016qwg}.

The fact that $\rho(s_F,s_B)\sim 1$ in the covered range of $\Delta \eta < 1.2$ and for all values of centrality 
indicates that the original sources in the early phase of the reaction may indeed be viewed as longitudinally extended objects (strings~\cite{Capella:1978rg}).
If such objects extend over rapidity in such a way that the $F$ and $B$ bins are always covered, then in each 
event $s_F=s_B$ and 
by definition we achieve the maximum correlation, $\rho(s_F,s_B)= 1$. 

The analysis presented in this paper was model-independent in the sense that we have only used the assumptions (a) and (b) from Sect.~\ref{sec:intro}, 
but have not referred to any specific model of the sources and particle production. With the method applied here and further spelled out in 
Refs.~\cite{Olszewski:2013qwa,Olszewski:2015xba}, such explicit models may be put to stringent tests with the help of 
experimental forward-backward correlation data.

\begin{appendix}

\section{Superposition model} \label{app:supp}

In this Appendix we recall the relevant formulas in the superposition model. A detailed 
derivation is presented in Ref.~\cite{Olszewski:2013qwa}.
Let the number of produced particles $n_A$ in bin $A$ ($A=F, B$) be composed of
independent emissions from $s_A$ sources, 
\begin{equation}
n_A=\sum_{i=1}^{s_A}m_i, \label{super}
\end{equation}
where $m_i$ is a random number of particles produced by the $i$th source. The distribution of $m_i$ is assumed to be universal, i.e., 
independent of the source $i$. Then one finds the well-known superposition formulas
\begin{eqnarray}
\langle n_A \rangle &=&  \langle m \rangle \langle s_A \rangle, \nonumber \\
{\rm var}(n_A) &=& {\rm var}(m) \langle s_A \rangle  + \langle m \rangle^2 {\rm var}(s_A) . \label{eq:var} 
\end{eqnarray}
Analogously, for the covariance between two well-separated bins we get immediately
\begin{eqnarray}
\langle n_F n_B \rangle = \langle \sum_{i=1}^{s_F} m_i  \sum_{j=1}^{s_B} m_j \rangle = 
\langle m \rangle^2 \langle s_F s_B \rangle, 
\end{eqnarray}
where we have used the fact that $\langle m_i m_j \rangle = 
\langle m \rangle^2$, holding for $i$ and $j$ belonging to two different well-separated bins.
As a result,
\begin{eqnarray}
{\rm cov}(n_F,n_B) = \langle m \rangle^2 {\rm cov}(s_F,s_B).  \label{eq:cors}
\end{eqnarray}

\section{Three stage approach}

Formulas (\ref{eq:var},\ref{eq:cors}) correspond to a single superposition step. 
In particular, such steps occur in the partonic phase, where partons are produced from the initial 
sources (strings), as well as in the late stage, where production of hadrons and their subsequent detection takes place. 
If superposition steps directly follow one another, the structure of Eqs.~(\ref{eq:var},\ref{eq:cors}) remains preserved.
For instance, this is the case of the hadron production step followed by the detection step (where the generic random variable $m_i$ would correspond 
to the detection of a hadron), hence we may combine these steps into a single one.
The intermediate evolution stage (hydrodynamics, transport) also preserves the structure of Eqs.~(\ref{eq:var},\ref{eq:cors}) and upon combining the three 
stages one finally has~\cite{Olszewski:2013qwa}
\begin{eqnarray}
\br{n_A} &=& \alpha \br{s_A}, \nonumber \\
\var{n_A} &=& \beta\br{s_A}+\gamma \var{s_A},\nonumber\\
\cov{n_F}{n_B} &=&  \gamma \cov{s_F}{s_B}.\label{eq:covvar}
\end{eqnarray}

Let $\mu$ denote the random number of partons produced in the first stage, and $m$ the random number of 
hadrons produced at final hadronization and registered by the detector. Further, if the number of partons 
is denoted with $p_A$ and the density of hydrodynamic fluid after the evolution as $h_A$, we may approximate 
the effect of the intermediate phase as
\begin{eqnarray}
h_A=t p_A, \label{eq:hyd}
\end{eqnarray}
where $t$ describes the intermediate evolution.\footnote{A more general affine variant of Eq.~(\ref{eq:hyd}) is used in Ref.~\cite{Olszewski:2013qwa}, but is 
does not affect the conclusions.}
As a result, we find
\begin{eqnarray}
\alpha &=& t \br{\mu} \br{m}, \nonumber \\
\beta  &=& t \br{\mu} \var{m} + t^2 \br{m}^2 \var{\mu}, \label{eq:abg} \\
\gamma &=& t^2 \br{\mu}^2 \br{m}^2. \nonumber
\end{eqnarray}

The inverse relations, relating moments of the sources via the moments on the measured hadrons, read
\begin{eqnarray}
\gamma \var{s_A} &=&  \var{n_A}-\delta \br{n_A},\nonumber\\
\gamma \cov{s_F}{s_B} &=& \cov{n_F}{n_B},\label{eq:covvar2}
\end{eqnarray}
where the $\delta$ parameter is given by relation
\begin{eqnarray}
\delta &=& \frac{\beta}{\alpha} = \omega(m) +t \br{m} \omega(\mu).\label{eq:delta}
\end{eqnarray}
Dividing Eqs.~(\ref{eq:covvar2}) side by side yields
\begin{eqnarray}
\rho\bn{s_F,s_B}&=&\frac{\cov{s_F}{s_B}}{\var{s_A}}=\frac{ \cov{s_F}{s_B}}{\var{n_A}-\delta\br{n_A}}=
\frac{\rho(n_F,n_B)}{1-\frac{\delta}{\omega\bn{n_A}}}, \label{eq:corrfinal}
\end{eqnarray}
which is our key formula~(\ref{eq:rhos}). Note that it involves only one combination of the parameters of the 
overlaid distributions and intermediate evolution, $\delta$.

The random variable $m$ in Eq.~(\ref{eq:delta}) corresponds to hadronization of the fluid folded with the detector acceptance. 
Due to its statistical nature, production of hadrons from the hydrodynamic fluid is well 
described by a Poisson distribution, whereas 
detector acceptance is modeled with a Bernoulli distribution. Folding of the Poisson and Bernoulli distributions yields a Poisson 
distribution, hence $\omega(m)=1$. 
Since all other parameters in Eq.~(\ref{eq:delta}) are positive, we conclude that 
\begin{eqnarray}
\delta > 1. \label{eq:cons}
\end{eqnarray}

Distributions of $\mu$ and $m$ are universal in the sense that they do not depend on the pseudorapidity of the bin or the centrality of the 
collision. The parameter $t$, which describes the hydrodynamic or transport response, is also expected to be approximately universal, meaning
linear response to the initial condition~\cite{Niemi:2012aj,Bzdak:2013rya,Bozek:2014cva,Fu:2015wba}.
Therefore we expect $\delta\simeq {\rm const}$.

\section{Multiple types of sources}

 \label{app:multi}

Our model uses one type of sources which emit particles $m$ with the same distribution, cf. Eq.~(\ref{super}). In this Appendix we show that 
under certain conditions our general results can be generalized to the case where we have more types of sources. For the simplest case of two 
kinds of sources
\begin{eqnarray}
n_A= \sum_{i=1}^{S_A} m_i + \sum_{i'=1}^{S'_A} m_{i'}, \;\;\; A=F,B.
\end{eqnarray}
Then, we find a generalization of Eq.~(\ref{eq:rhos}) in the form 
\begin{eqnarray}
\rho(u_F,u_B) &=& \frac{\rho(n_F,n_B)}
{1-\frac{\frac{\br{S_A}\var{m}+\br{S'_A}\var{m'}}
{\br{S_A}\br{m}+\br{S'_A}\br{m'}}}{\omega\bn{n_A}}},
\end{eqnarray}
where
\begin{eqnarray}
u_A=S_A\br{m}+S'_A\br{m'}. 
\end{eqnarray}
We note the same structure as in Eq.~(\ref{eq:rhos}), with $\delta$ replaced with the combination
\begin{eqnarray}
\delta=\frac{\br{S_A}\var{m}+\br{S'_A}\var{m'}}
{\br{S_A}\br{m}+\br{S'_A}\br{m'}}. \label{eq:newdel}
\end{eqnarray}
This combination is constant in two interesting cases: 
\begin{enumerate}
 \item $\br{S'_A}=\lambda \br{S_A}$,
 \item $\var{m}=\kappa \br{m}, \var{m'}=\kappa \br{m'}$,
\end{enumerate}
where constants $\lambda$ or $\kappa$ do not depend on centrality or the pseudorapidity separation. 
In the first case $\delta=(\var{m}+\lambda\var{m'})/(\br{m}+\lambda\br{m'})= {\rm const.}$, whereas 
in the second case $\delta=\kappa={\rm const.}$ 

The correlation $\rho(u_F,u_B)$ is a more complicated object which now plays the role of
$\rho(S_F,S_B)$ from Eq.~(\ref{eq:rhos},\ref{eq:omr1}). In a more general analysis with sources of multiple types we should keep it as is. 
A simplification occurs, however, when in each event 
$S'_A \simeq \lambda S_A$, i.e., the relative fluctuations are not too large. Then we have  
$\rho(u_F,u_B) \simeq \rho(S_F,S_B) \simeq \rho(S'_F,S'_B)$.

A physical realization of scenario~1) is the quark-diquark model of Ref.~\cite{Bialas:2007eg} for the A-A collisions, where we expect that 
(event-by-event) the numbers of wounded quarks and diquarks are proportional to each other.
Scenario 2)~occurs where the scaled variances of $m$ and $m'$ are equal. This is, e.g., the case of the Poisson distributions, or more general
negative binomial distributions with the same parameters controlling the scaled variance.

A generalization of the discussion of this Appendix to more than two types of sources is straightforward, with the sums 
showing up in the formulas extending from $2$ to $n$ kinds.

In conclusion, the analysis of this paper may be extended to the case where the superposition model involves more types of sources under the 
condition that the combination (\ref{eq:newdel}) is (approximately) constant. Conversely, the constant value of $\delta$ (as to a good approximation occurs 
in Fig.~(\ref{fig:deltacen})), does not require the assumption of a single type of sources.

\end{appendix}


\begin{thebibliography}{56}%
\makeatletter
\providecommand \@ifxundefined [1]{%
 \@ifx{#1\undefined}
}%
\providecommand \@ifnum [1]{%
 \ifnum #1\expandafter \@firstoftwo
 \else \expandafter \@secondoftwo
 \fi
}%
\providecommand \@ifx [1]{%
 \ifx #1\expandafter \@firstoftwo
 \else \expandafter \@secondoftwo
 \fi
}%
\providecommand \natexlab [1]{#1}%
\providecommand \enquote  [1]{``#1''}%
\providecommand \bibnamefont  [1]{#1}%
\providecommand \bibfnamefont [1]{#1}%
\providecommand \citenamefont [1]{#1}%
\providecommand \href@noop [0]{\@secondoftwo}%
\providecommand \href [0]{\begingroup \@sanitize@url \@href}%
\providecommand \@href[1]{\@@startlink{#1}\@@href}%
\providecommand \@@href[1]{\endgroup#1\@@endlink}%
\providecommand \@sanitize@url [0]{\catcode `\\12\catcode `\$12\catcode
  `\&12\catcode `\#12\catcode `\^12\catcode `\_12\catcode `\%12\relax}%
\providecommand \@@startlink[1]{}%
\providecommand \@@endlink[0]{}%
\providecommand \url  [0]{\begingroup\@sanitize@url \@url }%
\providecommand \@url [1]{\endgroup\@href {#1}{\urlprefix }}%
\providecommand \urlprefix  [0]{URL }%
\providecommand \Eprint [0]{\href }%
\providecommand \doibase [0]{http://dx.doi.org/}%
\providecommand \selectlanguage [0]{\@gobble}%
\providecommand \bibinfo  [0]{\@secondoftwo}%
\providecommand \bibfield  [0]{\@secondoftwo}%
\providecommand \translation [1]{[#1]}%
\providecommand \BibitemOpen [0]{}%
\providecommand \bibitemStop [0]{}%
\providecommand \bibitemNoStop [0]{.\EOS\space}%
\providecommand \EOS [0]{\spacefactor3000\relax}%
\providecommand \BibitemShut  [1]{\csname bibitem#1\endcsname}%
\let\auto@bib@innerbib\@empty
\bibitem [{\citenamefont {Sputowska}(2016)}]{Sputowska:2016xxx}%
  \BibitemOpen
  \bibfield  {author} {\bibinfo {author} {\bibfnamefont {I.}~\bibnamefont
  {Sputowska}},\ }\emph {\bibinfo {title}
  {\href{http://cds.cern.ch/record/2223886?ln=en}{Correlations in Particle
  Production in Nuclear Collisions at LHC Energies}}},\ \href
  {{http://cds.cern.ch/record/2223886?ln=en}} {Ph.D. thesis},\ \bibinfo
  {school} {{Institute of Nuclear Physics PAN, Cracow, Poland}}, \bibinfo
  {address} {{Cern Document Server}} (\bibinfo {year} {2016})\BibitemShut
  {NoStop}%
\bibitem [{\citenamefont {Olszewski}\ and\ \citenamefont
  {Broniowski}(2013)}]{Olszewski:2013qwa}%
  \BibitemOpen
  \bibfield  {author} {\bibinfo {author} {\bibfnamefont {A.}~\bibnamefont
  {Olszewski}}\ and\ \bibinfo {author} {\bibfnamefont {W.}~\bibnamefont
  {Broniowski}},\ }\href {\doibase 10.1103/PhysRevC.88.044913} {\bibfield
  {journal} {\bibinfo  {journal} {Phys.Rev.}\ }\textbf {\bibinfo {volume}
  {C88}},\ \bibinfo {pages} {044913} (\bibinfo {year} {2013})},\ \Eprint
  {http://arxiv.org/abs/1303.5280} {arXiv:1303.5280 [nucl-th]} \BibitemShut
  {NoStop}%
\bibitem [{\citenamefont {Olszewski}\ and\ \citenamefont
  {Broniowski}(2015)}]{Olszewski:2015xba}%
  \BibitemOpen
  \bibfield  {author} {\bibinfo {author} {\bibfnamefont {A.}~\bibnamefont
  {Olszewski}}\ and\ \bibinfo {author} {\bibfnamefont {W.}~\bibnamefont
  {Broniowski}},\ }\href {\doibase 10.1103/PhysRevC.92.024913} {\bibfield
  {journal} {\bibinfo  {journal} {Phys. Rev.}\ }\textbf {\bibinfo {volume}
  {C92}},\ \bibinfo {pages} {024913} (\bibinfo {year} {2015})},\ \Eprint
  {http://arxiv.org/abs/1502.05215} {arXiv:1502.05215 [nucl-th]} \BibitemShut
  {NoStop}%
\bibitem [{\citenamefont {Uhlig}\ \emph {et~al.}(1978)\citenamefont {Uhlig},
  \citenamefont {Derado}, \citenamefont {Meinke},\ and\ \citenamefont
  {Preissner}}]{Uhlig:1977dc}%
  \BibitemOpen
  \bibfield  {author} {\bibinfo {author} {\bibfnamefont {S.}~\bibnamefont
  {Uhlig}}, \bibinfo {author} {\bibfnamefont {I.}~\bibnamefont {Derado}},
  \bibinfo {author} {\bibfnamefont {R.}~\bibnamefont {Meinke}}, \ and\ \bibinfo
  {author} {\bibfnamefont {H.}~\bibnamefont {Preissner}},\ }\href {\doibase
  10.1016/0550-3213(78)90254-7} {\bibfield  {journal} {\bibinfo  {journal}
  {Nucl. Phys.}\ }\textbf {\bibinfo {volume} {B132}},\ \bibinfo {pages} {15}
  (\bibinfo {year} {1978})}\BibitemShut {NoStop}%
\bibitem [{\citenamefont {Alpgard}\ \emph {et~al.}(1983)\citenamefont {Alpgard}
  \emph {et~al.}}]{Alpgard:1983xp}%
  \BibitemOpen
  \bibfield  {author} {\bibinfo {author} {\bibfnamefont {K.}~\bibnamefont
  {Alpgard}} \emph {et~al.} (\bibinfo {collaboration} {UA5}),\ }\href {\doibase
  10.1016/0370-2693(83)91218-2} {\bibfield  {journal} {\bibinfo  {journal}
  {Phys. Lett.}\ }\textbf {\bibinfo {volume} {B123}},\ \bibinfo {pages} {361}
  (\bibinfo {year} {1983})}\BibitemShut {NoStop}%
\bibitem [{\citenamefont {Alner}\ \emph {et~al.}(1987)\citenamefont {Alner}
  \emph {et~al.}}]{Alner:1987wb}%
  \BibitemOpen
  \bibfield  {author} {\bibinfo {author} {\bibfnamefont {G.~J.}\ \bibnamefont
  {Alner}} \emph {et~al.} (\bibinfo {collaboration} {UA5}),\ }\href {\doibase
  10.1016/0370-1573(87)90130-X} {\bibfield  {journal} {\bibinfo  {journal}
  {Phys. Rept.}\ }\textbf {\bibinfo {volume} {154}},\ \bibinfo {pages} {247}
  (\bibinfo {year} {1987})}\BibitemShut {NoStop}%
\bibitem [{\citenamefont {Ansorge}\ \emph {et~al.}(1988)\citenamefont {Ansorge}
  \emph {et~al.}}]{Ansorge:1988fg}%
  \BibitemOpen
  \bibfield  {author} {\bibinfo {author} {\bibfnamefont {R.~E.}\ \bibnamefont
  {Ansorge}} \emph {et~al.} (\bibinfo {collaboration} {UA5}),\ }\href {\doibase
  10.1007/BF01579906} {\bibfield  {journal} {\bibinfo  {journal} {Z. Phys.}\
  }\textbf {\bibinfo {volume} {C37}},\ \bibinfo {pages} {191} (\bibinfo {year}
  {1988})}\BibitemShut {NoStop}%
\bibitem [{\citenamefont {Derado}\ \emph {et~al.}(1988)\citenamefont {Derado}
  \emph {et~al.}}]{Derado:1988ba}%
  \BibitemOpen
  \bibfield  {author} {\bibinfo {author} {\bibfnamefont {I.}~\bibnamefont
  {Derado}} \emph {et~al.},\ }\href {\doibase 10.1007/BF01559714} {\bibfield
  {journal} {\bibinfo  {journal} {Z. Phys.}\ }\textbf {\bibinfo {volume}
  {C40}},\ \bibinfo {pages} {25} (\bibinfo {year} {1988})}\BibitemShut
  {NoStop}%
\bibitem [{\citenamefont {Alexopoulos}\ \emph {et~al.}(1995)\citenamefont
  {Alexopoulos} \emph {et~al.}}]{Alexopoulos:1995ft}%
  \BibitemOpen
  \bibfield  {author} {\bibinfo {author} {\bibfnamefont {T.}~\bibnamefont
  {Alexopoulos}} \emph {et~al.} (\bibinfo {collaboration} {E735
  Collaboration}),\ }\href {\doibase 10.1016/0370-2693(95)00554-X} {\bibfield
  {journal} {\bibinfo  {journal} {Phys.Lett.}\ }\textbf {\bibinfo {volume}
  {B353}},\ \bibinfo {pages} {155} (\bibinfo {year} {1995})}\BibitemShut
  {NoStop}%
\bibitem [{\citenamefont {{B\"achler}}\ \emph {et~al.}(1992)\citenamefont
  {{B\"achler}} \emph {et~al.}}]{Bachler:1992psr}%
  \BibitemOpen
  \bibfield  {author} {\bibinfo {author} {\bibfnamefont {J.}~\bibnamefont
  {{B\"achler}}} \emph {et~al.} (\bibinfo {collaboration} {NA35}),\ }\href
  {\doibase 10.1007/BF01565942} {\bibfield  {journal} {\bibinfo  {journal} {Z.
  Phys.}\ }\textbf {\bibinfo {volume} {C56}},\ \bibinfo {pages} {347} (\bibinfo
  {year} {1992})}\BibitemShut {NoStop}%
\bibitem [{\citenamefont {Akiba}\ \emph {et~al.}(1997)\citenamefont {Akiba}
  \emph {et~al.}}]{Akiba:1997yg}%
  \BibitemOpen
  \bibfield  {author} {\bibinfo {author} {\bibfnamefont {Y.}~\bibnamefont
  {Akiba}} \emph {et~al.} (\bibinfo {collaboration} {E802}),\ }\href {\doibase
  10.1103/PhysRevC.56.1544} {\bibfield  {journal} {\bibinfo  {journal} {Phys.
  Rev.}\ }\textbf {\bibinfo {volume} {C56}},\ \bibinfo {pages} {1544} (\bibinfo
  {year} {1997})}\BibitemShut {NoStop}%
\bibitem [{\citenamefont {Back}\ \emph {et~al.}(2006)\citenamefont {Back} \emph
  {et~al.}}]{Back:2006id}%
  \BibitemOpen
  \bibfield  {author} {\bibinfo {author} {\bibfnamefont {B.~B.}\ \bibnamefont
  {Back}} \emph {et~al.} (\bibinfo {collaboration} {PHOBOS}),\ }\href {\doibase
  10.1103/PhysRevC.74.011901} {\bibfield  {journal} {\bibinfo  {journal} {Phys.
  Rev.}\ }\textbf {\bibinfo {volume} {C74}},\ \bibinfo {pages} {011901}
  (\bibinfo {year} {2006})},\ \Eprint {http://arxiv.org/abs/nucl-ex/0603026}
  {arXiv:nucl-ex/0603026 [nucl-ex]} \BibitemShut {NoStop}%
\bibitem [{\citenamefont {Abelev}\ \emph {et~al.}(2009)\citenamefont {Abelev}
  \emph {et~al.}}]{Abelev:2009ag}%
  \BibitemOpen
  \bibfield  {author} {\bibinfo {author} {\bibfnamefont {B.}~\bibnamefont
  {Abelev}} \emph {et~al.} (\bibinfo {collaboration} {STAR Collaboration}),\
  }\href {\doibase 10.1103/PhysRevLett.103.172301} {\bibfield  {journal}
  {\bibinfo  {journal} {Phys.Rev.Lett.}\ }\textbf {\bibinfo {volume} {103}},\
  \bibinfo {pages} {172301} (\bibinfo {year} {2009})},\ \Eprint
  {http://arxiv.org/abs/0905.0237} {arXiv:0905.0237 [nucl-ex]} \BibitemShut
  {NoStop}%
\bibitem [{\citenamefont {Tarnowsky}(2010)}]{Tarnowsky:2010qp}%
  \BibitemOpen
  \bibfield  {author} {\bibinfo {author} {\bibfnamefont {T.~J.}\ \bibnamefont
  {Tarnowsky}},\ }\href {\doibase 10.1088/1742-6596/230/1/012025} {\bibfield
  {journal} {\bibinfo  {journal} {J.Phys.Conf.Ser.}\ }\textbf {\bibinfo
  {volume} {230}},\ \bibinfo {pages} {012025} (\bibinfo {year} {2010})},\
  \Eprint {http://arxiv.org/abs/1005.1895} {arXiv:1005.1895 [nucl-ex]}
  \BibitemShut {NoStop}%
\bibitem [{\citenamefont {Aad}\ \emph {et~al.}(2012)\citenamefont {Aad} \emph
  {et~al.}}]{ATLAS:2012as}%
  \BibitemOpen
  \bibfield  {author} {\bibinfo {author} {\bibfnamefont {G.}~\bibnamefont
  {Aad}} \emph {et~al.} (\bibinfo {collaboration} {ATLAS}),\ }\href {\doibase
  10.1007/JHEP07(2012)019} {\bibfield  {journal} {\bibinfo  {journal} {JHEP}\
  }\textbf {\bibinfo {volume} {07}},\ \bibinfo {pages} {019} (\bibinfo {year}
  {2012})},\ \Eprint {http://arxiv.org/abs/1203.3100} {arXiv:1203.3100
  [hep-ex]} \BibitemShut {NoStop}%
\bibitem [{\citenamefont {Jia}\ \emph {et~al.}(2016)\citenamefont {Jia},
  \citenamefont {Radhakrishnan},\ and\ \citenamefont {Zhou}}]{Jia:2015jga}%
  \BibitemOpen
  \bibfield  {author} {\bibinfo {author} {\bibfnamefont {J.}~\bibnamefont
  {Jia}}, \bibinfo {author} {\bibfnamefont {S.}~\bibnamefont {Radhakrishnan}},
  \ and\ \bibinfo {author} {\bibfnamefont {M.}~\bibnamefont {Zhou}},\ }\href
  {\doibase 10.1103/PhysRevC.93.044905} {\bibfield  {journal} {\bibinfo
  {journal} {Phys. Rev.}\ }\textbf {\bibinfo {volume} {C93}},\ \bibinfo {pages}
  {044905} (\bibinfo {year} {2016})},\ \Eprint
  {http://arxiv.org/abs/1506.03496} {arXiv:1506.03496 [nucl-th]} \BibitemShut
  {NoStop}%
\bibitem [{\citenamefont {Adam}\ \emph {et~al.}(2015)\citenamefont {Adam} \emph
  {et~al.}}]{Adam:2015mya}%
  \BibitemOpen
  \bibfield  {author} {\bibinfo {author} {\bibfnamefont {J.}~\bibnamefont
  {Adam}} \emph {et~al.} (\bibinfo {collaboration} {ALICE}),\ }\href {\doibase
  10.1007/JHEP05(2015)097} {\bibfield  {journal} {\bibinfo  {journal} {JHEP}\
  }\textbf {\bibinfo {volume} {05}},\ \bibinfo {pages} {097} (\bibinfo {year}
  {2015})},\ \Eprint {http://arxiv.org/abs/1502.00230} {arXiv:1502.00230
  [nucl-ex]} \BibitemShut {NoStop}%
\bibitem [{\citenamefont {Aaboud}\ \emph {et~al.}(2016)\citenamefont {Aaboud}
  \emph {et~al.}}]{Aaboud:2016jnr}%
  \BibitemOpen
  \bibfield  {author} {\bibinfo {author} {\bibfnamefont {M.}~\bibnamefont
  {Aaboud}} \emph {et~al.} (\bibinfo {collaboration} {ATLAS}),\ }\href@noop {}
  {\  (\bibinfo {year} {2016})},\ \Eprint {http://arxiv.org/abs/1606.08170}
  {arXiv:1606.08170 [hep-ex]} \BibitemShut {NoStop}%
\bibitem [{\citenamefont {Capella}\ and\ \citenamefont
  {Krzywicki}(1978)}]{Capella:1978rg}%
  \BibitemOpen
  \bibfield  {author} {\bibinfo {author} {\bibfnamefont {A.}~\bibnamefont
  {Capella}}\ and\ \bibinfo {author} {\bibfnamefont {A.}~\bibnamefont
  {Krzywicki}},\ }\href {\doibase 10.1103/PhysRevD.18.4120} {\bibfield
  {journal} {\bibinfo  {journal} {Phys.Rev.}\ }\textbf {\bibinfo {volume}
  {D18}},\ \bibinfo {pages} {4120} (\bibinfo {year} {1978})}\BibitemShut
  {NoStop}%
\bibitem [{\citenamefont {Kaidalov}\ and\ \citenamefont
  {Ter-Martirosian}(1982)}]{Kaidalov:1982xe}%
  \BibitemOpen
  \bibfield  {author} {\bibinfo {author} {\bibfnamefont {A.~B.}\ \bibnamefont
  {Kaidalov}}\ and\ \bibinfo {author} {\bibfnamefont {K.~A.}\ \bibnamefont
  {Ter-Martirosian}},\ }\href {\doibase 10.1016/0370-2693(82)90556-1}
  {\bibfield  {journal} {\bibinfo  {journal} {Phys. Lett.}\ }\textbf {\bibinfo
  {volume} {B117}},\ \bibinfo {pages} {247} (\bibinfo {year}
  {1982})}\BibitemShut {NoStop}%
\bibitem [{\citenamefont {Chou}\ and\ \citenamefont
  {Yang}(1984)}]{Chou:1984wp}%
  \BibitemOpen
  \bibfield  {author} {\bibinfo {author} {\bibfnamefont {T.~T.}\ \bibnamefont
  {Chou}}\ and\ \bibinfo {author} {\bibfnamefont {C.~N.}\ \bibnamefont
  {Yang}},\ }\href {\doibase 10.1016/0370-2693(84)90478-7} {\bibfield
  {journal} {\bibinfo  {journal} {Phys. Lett.}\ }\textbf {\bibinfo {volume}
  {B135}},\ \bibinfo {pages} {175} (\bibinfo {year} {1984})}\BibitemShut
  {NoStop}%
\bibitem [{\citenamefont {Capella}\ \emph {et~al.}(1994)\citenamefont
  {Capella}, \citenamefont {Sukhatme}, \citenamefont {Tan},\ and\ \citenamefont
  {Tran Thanh~Van}}]{Capella:1992yb}%
  \BibitemOpen
  \bibfield  {author} {\bibinfo {author} {\bibfnamefont {A.}~\bibnamefont
  {Capella}}, \bibinfo {author} {\bibfnamefont {U.}~\bibnamefont {Sukhatme}},
  \bibinfo {author} {\bibfnamefont {C.-I.}\ \bibnamefont {Tan}}, \ and\
  \bibinfo {author} {\bibfnamefont {J.}~\bibnamefont {Tran Thanh~Van}},\ }\href
  {\doibase 10.1016/0370-1573(94)90064-7} {\bibfield  {journal} {\bibinfo
  {journal} {Phys. Rept.}\ }\textbf {\bibinfo {volume} {236}},\ \bibinfo
  {pages} {225} (\bibinfo {year} {1994})}\BibitemShut {NoStop}%
\bibitem [{\citenamefont {Amelin}\ \emph {et~al.}(1994)\citenamefont {Amelin},
  \citenamefont {Armesto}, \citenamefont {Braun}, \citenamefont {Ferreiro},\
  and\ \citenamefont {Pajares}}]{Amelin:1994mf}%
  \BibitemOpen
  \bibfield  {author} {\bibinfo {author} {\bibfnamefont {N.~S.}\ \bibnamefont
  {Amelin}}, \bibinfo {author} {\bibfnamefont {N.}~\bibnamefont {Armesto}},
  \bibinfo {author} {\bibfnamefont {M.~A.}\ \bibnamefont {Braun}}, \bibinfo
  {author} {\bibfnamefont {E.~G.}\ \bibnamefont {Ferreiro}}, \ and\ \bibinfo
  {author} {\bibfnamefont {C.}~\bibnamefont {Pajares}},\ }\href {\doibase
  10.1103/PhysRevLett.73.2813} {\bibfield  {journal} {\bibinfo  {journal}
  {Phys.Rev.Lett.}\ }\textbf {\bibinfo {volume} {73}},\ \bibinfo {pages} {2813}
  (\bibinfo {year} {1994})}\BibitemShut {NoStop}%
\bibitem [{\citenamefont {Braun}\ \emph {et~al.}(2000)\citenamefont {Braun},
  \citenamefont {Pajares},\ and\ \citenamefont {Vechernin}}]{Braun:2000cc}%
  \BibitemOpen
  \bibfield  {author} {\bibinfo {author} {\bibfnamefont {M.}~\bibnamefont
  {Braun}}, \bibinfo {author} {\bibfnamefont {C.}~\bibnamefont {Pajares}}, \
  and\ \bibinfo {author} {\bibfnamefont {V.}~\bibnamefont {Vechernin}},\ }\href
  {\doibase 10.1016/S0370-2693(00)01127-8} {\bibfield  {journal} {\bibinfo
  {journal} {Phys.Lett.}\ }\textbf {\bibinfo {volume} {B493}},\ \bibinfo
  {pages} {54} (\bibinfo {year} {2000})},\ \Eprint
  {http://arxiv.org/abs/hep-ph/0007241} {arXiv:hep-ph/0007241 [hep-ph]}
  \BibitemShut {NoStop}%
\bibitem [{\citenamefont {Giovannini}\ and\ \citenamefont
  {Ugoccioni}(2002)}]{Giovannini:2002za}%
  \BibitemOpen
  \bibfield  {author} {\bibinfo {author} {\bibfnamefont {A.}~\bibnamefont
  {Giovannini}}\ and\ \bibinfo {author} {\bibfnamefont {R.}~\bibnamefont
  {Ugoccioni}},\ }\href {\doibase 10.1103/PhysRevD.66.034001} {\bibfield
  {journal} {\bibinfo  {journal} {Phys. Rev.}\ }\textbf {\bibinfo {volume}
  {D66}},\ \bibinfo {pages} {034001} (\bibinfo {year} {2002})},\ \Eprint
  {http://arxiv.org/abs/hep-ph/0205156} {arXiv:hep-ph/0205156 [hep-ph]}
  \BibitemShut {NoStop}%
\bibitem [{\citenamefont {Braun}\ \emph {et~al.}(2004)\citenamefont {Braun},
  \citenamefont {Kolevatov}, \citenamefont {Pajares},\ and\ \citenamefont
  {Vechernin}}]{Braun:2003fn}%
  \BibitemOpen
  \bibfield  {author} {\bibinfo {author} {\bibfnamefont {M.}~\bibnamefont
  {Braun}}, \bibinfo {author} {\bibfnamefont {R.}~\bibnamefont {Kolevatov}},
  \bibinfo {author} {\bibfnamefont {C.}~\bibnamefont {Pajares}}, \ and\
  \bibinfo {author} {\bibfnamefont {V.}~\bibnamefont {Vechernin}},\ }\href
  {\doibase 10.1140/epjc/s2003-01443-6} {\bibfield  {journal} {\bibinfo
  {journal} {Eur.Phys.J.}\ }\textbf {\bibinfo {volume} {C32}},\ \bibinfo
  {pages} {535} (\bibinfo {year} {2004})},\ \Eprint
  {http://arxiv.org/abs/hep-ph/0307056} {arXiv:hep-ph/0307056 [hep-ph]}
  \BibitemShut {NoStop}%
\bibitem [{\citenamefont {Brogueira}\ \emph {et~al.}(2007)\citenamefont
  {Brogueira}, \citenamefont {Dias~de Deus},\ and\ \citenamefont
  {Milhano}}]{Brogueira:2007ub}%
  \BibitemOpen
  \bibfield  {author} {\bibinfo {author} {\bibfnamefont {P.}~\bibnamefont
  {Brogueira}}, \bibinfo {author} {\bibfnamefont {J.}~\bibnamefont {Dias~de
  Deus}}, \ and\ \bibinfo {author} {\bibfnamefont {J.~G.}\ \bibnamefont
  {Milhano}},\ }\href {\doibase 10.1103/PhysRevC.76.064901} {\bibfield
  {journal} {\bibinfo  {journal} {Phys. Rev.}\ }\textbf {\bibinfo {volume}
  {C76}},\ \bibinfo {pages} {064901} (\bibinfo {year} {2007})},\ \Eprint
  {http://arxiv.org/abs/0709.3913} {arXiv:0709.3913 [hep-ph]} \BibitemShut
  {NoStop}%
\bibitem [{\citenamefont {Armesto}\ \emph
  {et~al.}(2007{\natexlab{a}})\citenamefont {Armesto}, \citenamefont {Braun},\
  and\ \citenamefont {Pajares}}]{Armesto:2007ia}%
  \BibitemOpen
  \bibfield  {author} {\bibinfo {author} {\bibfnamefont {N.}~\bibnamefont
  {Armesto}}, \bibinfo {author} {\bibfnamefont {M.}~\bibnamefont {Braun}}, \
  and\ \bibinfo {author} {\bibfnamefont {C.}~\bibnamefont {Pajares}},\ }\href
  {\doibase 10.1103/PhysRevC.75.054902} {\bibfield  {journal} {\bibinfo
  {journal} {Phys.Rev.}\ }\textbf {\bibinfo {volume} {C75}},\ \bibinfo {pages}
  {054902} (\bibinfo {year} {2007}{\natexlab{a}})},\ \Eprint
  {http://arxiv.org/abs/hep-ph/0702216} {arXiv:hep-ph/0702216 [HEP-PH]}
  \BibitemShut {NoStop}%
\bibitem [{\citenamefont {Armesto}\ \emph
  {et~al.}(2007{\natexlab{b}})\citenamefont {Armesto}, \citenamefont
  {McLerran},\ and\ \citenamefont {Pajares}}]{Armesto:2006bv}%
  \BibitemOpen
  \bibfield  {author} {\bibinfo {author} {\bibfnamefont {N.}~\bibnamefont
  {Armesto}}, \bibinfo {author} {\bibfnamefont {L.}~\bibnamefont {McLerran}}, \
  and\ \bibinfo {author} {\bibfnamefont {C.}~\bibnamefont {Pajares}},\ }\href
  {\doibase 10.1016/j.nuclphysa.2006.10.074} {\bibfield  {journal} {\bibinfo
  {journal} {Nucl. Phys.}\ }\textbf {\bibinfo {volume} {A781}},\ \bibinfo
  {pages} {201} (\bibinfo {year} {2007}{\natexlab{b}})},\ \Eprint
  {http://arxiv.org/abs/hep-ph/0607345} {arXiv:hep-ph/0607345} \BibitemShut
  {NoStop}%
\bibitem [{\citenamefont {Vechernin}\ and\ \citenamefont
  {Kolevatov}(2007)}]{Vechernin:2007zza}%
  \BibitemOpen
  \bibfield  {author} {\bibinfo {author} {\bibfnamefont {V.}~\bibnamefont
  {Vechernin}}\ and\ \bibinfo {author} {\bibfnamefont {R.}~\bibnamefont
  {Kolevatov}},\ }\href {\doibase 10.1134/S1063778807100158} {\bibfield
  {journal} {\bibinfo  {journal} {Phys.Atom.Nucl.}\ }\textbf {\bibinfo {volume}
  {70}},\ \bibinfo {pages} {1797} (\bibinfo {year} {2007})}\BibitemShut
  {NoStop}%
\bibitem [{\citenamefont {Braun}(2008)}]{Braun:2007rf}%
  \BibitemOpen
  \bibfield  {author} {\bibinfo {author} {\bibfnamefont {M.}~\bibnamefont
  {Braun}},\ }\href {\doibase 10.1016/j.nuclphysa.2008.02.301} {\bibfield
  {journal} {\bibinfo  {journal} {Nucl.Phys.}\ }\textbf {\bibinfo {volume}
  {A806}},\ \bibinfo {pages} {230} (\bibinfo {year} {2008})},\ \Eprint
  {http://arxiv.org/abs/0711.3268} {arXiv:0711.3268 [hep-ph]} \BibitemShut
  {NoStop}%
\bibitem [{\citenamefont {Konchakovski}\ \emph {et~al.}(2009)\citenamefont
  {Konchakovski}, \citenamefont {Hauer}, \citenamefont {Torrieri},
  \citenamefont {Gorenstein},\ and\ \citenamefont
  {Bratkovskaya}}]{Konchakovski:2008cf}%
  \BibitemOpen
  \bibfield  {author} {\bibinfo {author} {\bibfnamefont {V.~P.}\ \bibnamefont
  {Konchakovski}}, \bibinfo {author} {\bibfnamefont {M.}~\bibnamefont {Hauer}},
  \bibinfo {author} {\bibfnamefont {G.}~\bibnamefont {Torrieri}}, \bibinfo
  {author} {\bibfnamefont {M.~I.}\ \bibnamefont {Gorenstein}}, \ and\ \bibinfo
  {author} {\bibfnamefont {E.~L.}\ \bibnamefont {Bratkovskaya}},\ }\href
  {\doibase 10.1103/PhysRevC.79.034910} {\bibfield  {journal} {\bibinfo
  {journal} {Phys. Rev.}\ }\textbf {\bibinfo {volume} {C79}},\ \bibinfo {pages}
  {034910} (\bibinfo {year} {2009})},\ \Eprint {http://arxiv.org/abs/0812.3967}
  {arXiv:0812.3967 [nucl-th]} \BibitemShut {NoStop}%
\bibitem [{\citenamefont {Bzdak}\ and\ \citenamefont
  {Wo\'zniak}(2010)}]{Bzdak:2009dr}%
  \BibitemOpen
  \bibfield  {author} {\bibinfo {author} {\bibfnamefont {A.}~\bibnamefont
  {Bzdak}}\ and\ \bibinfo {author} {\bibfnamefont {K.}~\bibnamefont
  {Wo\'zniak}},\ }\href {\doibase 10.1103/PhysRevC.81.034908} {\bibfield
  {journal} {\bibinfo  {journal} {Phys. Rev.}\ }\textbf {\bibinfo {volume}
  {C81}},\ \bibinfo {pages} {034908} (\bibinfo {year} {2010})},\ \Eprint
  {http://arxiv.org/abs/0911.4696} {arXiv:0911.4696 [hep-ph]} \BibitemShut
  {NoStop}%
\bibitem [{\citenamefont {Lappi}\ and\ \citenamefont
  {McLerran}(2010)}]{Lappi:2009vb}%
  \BibitemOpen
  \bibfield  {author} {\bibinfo {author} {\bibfnamefont {T.}~\bibnamefont
  {Lappi}}\ and\ \bibinfo {author} {\bibfnamefont {L.}~\bibnamefont
  {McLerran}},\ }\href {\doibase 10.1016/j.nuclphysa.2009.11.003} {\bibfield
  {journal} {\bibinfo  {journal} {Nucl.Phys.}\ }\textbf {\bibinfo {volume}
  {A832}},\ \bibinfo {pages} {330} (\bibinfo {year} {2010})},\ \Eprint
  {http://arxiv.org/abs/0909.0428} {arXiv:0909.0428 [hep-ph]} \BibitemShut
  {NoStop}%
\bibitem [{\citenamefont {Bo\.zek}\ \emph {et~al.}(2011)\citenamefont
  {Bo\.zek}, \citenamefont {Broniowski},\ and\ \citenamefont
  {Moreira}}]{Bozek:2010vz}%
  \BibitemOpen
  \bibfield  {author} {\bibinfo {author} {\bibfnamefont {P.}~\bibnamefont
  {Bo\.zek}}, \bibinfo {author} {\bibfnamefont {W.}~\bibnamefont {Broniowski}},
  \ and\ \bibinfo {author} {\bibfnamefont {J.}~\bibnamefont {Moreira}},\ }\href
  {\doibase 10.1103/PhysRevC.83.034911} {\bibfield  {journal} {\bibinfo
  {journal} {Phys. Rev.}\ }\textbf {\bibinfo {volume} {C83}},\ \bibinfo {pages}
  {034911} (\bibinfo {year} {2011})},\ \Eprint {http://arxiv.org/abs/1011.3354}
  {arXiv:1011.3354 [nucl-th]} \BibitemShut {NoStop}%
\bibitem [{\citenamefont {Dias~de Deus}\ and\ \citenamefont
  {Pajares}(2011)}]{deDeus:2010id}%
  \BibitemOpen
  \bibfield  {author} {\bibinfo {author} {\bibfnamefont {J.}~\bibnamefont
  {Dias~de Deus}}\ and\ \bibinfo {author} {\bibfnamefont {C.}~\bibnamefont
  {Pajares}},\ }\href {\doibase 10.1016/j.physletb.2010.11.017} {\bibfield
  {journal} {\bibinfo  {journal} {Phys. Lett.}\ }\textbf {\bibinfo {volume}
  {B695}},\ \bibinfo {pages} {211} (\bibinfo {year} {2011})},\ \Eprint
  {http://arxiv.org/abs/1011.1099} {arXiv:1011.1099 [hep-ph]} \BibitemShut
  {NoStop}%
\bibitem [{\citenamefont {Bia\l{}as}\ and\ \citenamefont
  {Zalewski}(2011{\natexlab{a}})}]{Bialas:2011xk}%
  \BibitemOpen
  \bibfield  {author} {\bibinfo {author} {\bibfnamefont {A.}~\bibnamefont
  {Bia\l{}as}}\ and\ \bibinfo {author} {\bibfnamefont {K.}~\bibnamefont
  {Zalewski}},\ }\href {\doibase 10.1016/j.nuclphysa.2011.05.006} {\bibfield
  {journal} {\bibinfo  {journal} {Nucl.Phys.}\ }\textbf {\bibinfo {volume}
  {A860}},\ \bibinfo {pages} {56} (\bibinfo {year} {2011}{\natexlab{a}})},\
  \Eprint {http://arxiv.org/abs/1101.1907} {arXiv:1101.1907 [hep-ph]}
  \BibitemShut {NoStop}%
\bibitem [{\citenamefont {Bia\l{}as}\ and\ \citenamefont
  {Zalewski}(2011{\natexlab{b}})}]{Bialas:2011vj}%
  \BibitemOpen
  \bibfield  {author} {\bibinfo {author} {\bibfnamefont {A.}~\bibnamefont
  {Bia\l{}as}}\ and\ \bibinfo {author} {\bibfnamefont {K.}~\bibnamefont
  {Zalewski}},\ }\href {\doibase 10.1016/j.physletb.2011.03.036} {\bibfield
  {journal} {\bibinfo  {journal} {Phys. Lett.}\ }\textbf {\bibinfo {volume}
  {B698}},\ \bibinfo {pages} {416} (\bibinfo {year} {2011}{\natexlab{b}})},\
  \Eprint {http://arxiv.org/abs/1101.5706} {arXiv:1101.5706 [hep-ph]}
  \BibitemShut {NoStop}%
\bibitem [{\citenamefont {Bzdak}(2012)}]{Bzdak:2011nb}%
  \BibitemOpen
  \bibfield  {author} {\bibinfo {author} {\bibfnamefont {A.}~\bibnamefont
  {Bzdak}},\ }\href {\doibase 10.1103/PhysRevC.85.051901} {\bibfield  {journal}
  {\bibinfo  {journal} {Phys.Rev.}\ }\textbf {\bibinfo {volume} {C85}},\
  \bibinfo {pages} {051901} (\bibinfo {year} {2012})},\ \Eprint
  {http://arxiv.org/abs/1108.0882} {arXiv:1108.0882 [hep-ph]} \BibitemShut
  {NoStop}%
\bibitem [{\citenamefont {Bzdak}\ and\ \citenamefont
  {Teaney}(2013)}]{Bzdak:2012tp}%
  \BibitemOpen
  \bibfield  {author} {\bibinfo {author} {\bibfnamefont {A.}~\bibnamefont
  {Bzdak}}\ and\ \bibinfo {author} {\bibfnamefont {D.}~\bibnamefont {Teaney}},\
  }\href {\doibase 10.1103/PhysRevC.87.024906} {\bibfield  {journal} {\bibinfo
  {journal} {Phys.Rev.}\ }\textbf {\bibinfo {volume} {C87}},\ \bibinfo {pages}
  {024906} (\bibinfo {year} {2013})},\ \Eprint {http://arxiv.org/abs/1210.1965}
  {arXiv:1210.1965 [nucl-th]} \BibitemShut {NoStop}%
\bibitem [{\citenamefont {Vechernin}(2012)}]{Vechernin:2012bz}%
  \BibitemOpen
  \bibfield  {author} {\bibinfo {author} {\bibfnamefont {V.~V.}\ \bibnamefont
  {Vechernin}},\ }\href@noop {} {\  (\bibinfo {year} {2012})},\ \Eprint
  {http://arxiv.org/abs/1210.7588} {arXiv:1210.7588 [hep-ph]} \BibitemShut
  {NoStop}%
\bibitem [{\citenamefont {Bialas}\ \emph {et~al.}(2013)\citenamefont {Bialas},
  \citenamefont {Bzdak},\ and\ \citenamefont {Zalewski}}]{Bialas:2013xea}%
  \BibitemOpen
  \bibfield  {author} {\bibinfo {author} {\bibfnamefont {A.}~\bibnamefont
  {Bialas}}, \bibinfo {author} {\bibfnamefont {A.}~\bibnamefont {Bzdak}}, \
  and\ \bibinfo {author} {\bibfnamefont {K.}~\bibnamefont {Zalewski}},\ }\href
  {\doibase 10.5506/APhysPolBSupp.6.463} {\bibfield  {journal} {\bibinfo
  {journal} {Acta Phys.Polon.Supp.}\ }\textbf {\bibinfo {volume} {6}},\
  \bibinfo {pages} {463} (\bibinfo {year} {2013})}\BibitemShut {NoStop}%
\bibitem [{\citenamefont {De}\ \emph {et~al.}(2013)\citenamefont {De},
  \citenamefont {Tarnowsky}, \citenamefont {Nayak}, \citenamefont
  {Scharenberg},\ and\ \citenamefont {Srivastava}}]{De:2013bta}%
  \BibitemOpen
  \bibfield  {author} {\bibinfo {author} {\bibfnamefont {S.}~\bibnamefont
  {De}}, \bibinfo {author} {\bibfnamefont {T.}~\bibnamefont {Tarnowsky}},
  \bibinfo {author} {\bibfnamefont {T.~K.}\ \bibnamefont {Nayak}}, \bibinfo
  {author} {\bibfnamefont {R.~P.}\ \bibnamefont {Scharenberg}}, \ and\ \bibinfo
  {author} {\bibfnamefont {B.~K.}\ \bibnamefont {Srivastava}},\ }\href
  {\doibase 10.1103/PhysRevC.88.044903} {\bibfield  {journal} {\bibinfo
  {journal} {Phys. Rev.}\ }\textbf {\bibinfo {volume} {C88}},\ \bibinfo {pages}
  {044903} (\bibinfo {year} {2013})},\ \Eprint {http://arxiv.org/abs/1309.7242}
  {arXiv:1309.7242 [nucl-ex]} \BibitemShut {NoStop}%
\bibitem [{\citenamefont {Ma}\ and\ \citenamefont {Bzdak}(2014)}]{Ma:2014pva}%
  \BibitemOpen
  \bibfield  {author} {\bibinfo {author} {\bibfnamefont {G.-L.}\ \bibnamefont
  {Ma}}\ and\ \bibinfo {author} {\bibfnamefont {A.}~\bibnamefont {Bzdak}},\
  }\href {\doibase 10.1016/j.physletb.2014.10.066} {\bibfield  {journal}
  {\bibinfo  {journal} {Phys.Lett.}\ }\textbf {\bibinfo {volume} {B739}},\
  \bibinfo {pages} {209} (\bibinfo {year} {2014})},\ \Eprint
  {http://arxiv.org/abs/1404.4129} {arXiv:1404.4129 [hep-ph]} \BibitemShut
  {NoStop}%
\bibitem [{\citenamefont {Bzdak}\ and\ \citenamefont
  {Bozek}(2016)}]{Bzdak:2015dja}%
  \BibitemOpen
  \bibfield  {author} {\bibinfo {author} {\bibfnamefont {A.}~\bibnamefont
  {Bzdak}}\ and\ \bibinfo {author} {\bibfnamefont {P.}~\bibnamefont {Bozek}},\
  }\href {\doibase 10.1103/PhysRevC.93.024903} {\bibfield  {journal} {\bibinfo
  {journal} {Phys. Rev.}\ }\textbf {\bibinfo {volume} {C93}},\ \bibinfo {pages}
  {024903} (\bibinfo {year} {2016})},\ \Eprint
  {http://arxiv.org/abs/1509.02967} {arXiv:1509.02967 [hep-ph]} \BibitemShut
  {NoStop}%
\bibitem [{\citenamefont {Bzdak}\ and\ \citenamefont
  {Dusling}(2016)}]{Bzdak:2015eii}%
  \BibitemOpen
  \bibfield  {author} {\bibinfo {author} {\bibfnamefont {A.}~\bibnamefont
  {Bzdak}}\ and\ \bibinfo {author} {\bibfnamefont {K.}~\bibnamefont
  {Dusling}},\ }\href {\doibase 10.1103/PhysRevC.93.031901} {\bibfield
  {journal} {\bibinfo  {journal} {Phys. Rev.}\ }\textbf {\bibinfo {volume}
  {C93}},\ \bibinfo {pages} {031901} (\bibinfo {year} {2016})},\ \Eprint
  {http://arxiv.org/abs/1511.03620} {arXiv:1511.03620 [hep-ph]} \BibitemShut
  {NoStop}%
\bibitem [{\citenamefont {Vechernin}(2015)}]{Vechernin:2015upa}%
  \BibitemOpen
  \bibfield  {author} {\bibinfo {author} {\bibfnamefont {V.}~\bibnamefont
  {Vechernin}},\ }\href {\doibase 10.1016/j.nuclphysa.2015.03.009} {\bibfield
  {journal} {\bibinfo  {journal} {Nucl. Phys.}\ }\textbf {\bibinfo {volume}
  {A939}},\ \bibinfo {pages} {21} (\bibinfo {year} {2015})}\BibitemShut
  {NoStop}%
\bibitem [{\citenamefont {Kharzeev}\ and\ \citenamefont
  {Nardi}(2001)}]{Kharzeev:2000ph}%
  \BibitemOpen
  \bibfield  {author} {\bibinfo {author} {\bibfnamefont {D.}~\bibnamefont
  {Kharzeev}}\ and\ \bibinfo {author} {\bibfnamefont {M.}~\bibnamefont
  {Nardi}},\ }\href {\doibase 10.1016/S0370-2693(01)00457-9} {\bibfield
  {journal} {\bibinfo  {journal} {Phys. Lett.}\ }\textbf {\bibinfo {volume}
  {B507}},\ \bibinfo {pages} {121} (\bibinfo {year} {2001})},\ \Eprint
  {http://arxiv.org/abs/nucl-th/0012025} {arXiv:nucl-th/0012025} \BibitemShut
  {NoStop}%
\bibitem [{\citenamefont {Bia\l{}as}\ \emph {et~al.}(1976)\citenamefont
  {Bia\l{}as}, \citenamefont {B\l{}eszy\'nski},\ and\ \citenamefont
  {Czy\.z}}]{Bialas:1976ed}%
  \BibitemOpen
  \bibfield  {author} {\bibinfo {author} {\bibfnamefont {A.}~\bibnamefont
  {Bia\l{}as}}, \bibinfo {author} {\bibfnamefont {M.}~\bibnamefont
  {B\l{}eszy\'nski}}, \ and\ \bibinfo {author} {\bibfnamefont {W.}~\bibnamefont
  {Czy\.z}},\ }\href@noop {} {\bibfield  {journal} {\bibinfo  {journal} {Nucl.
  Phys.}\ }\textbf {\bibinfo {volume} {B111}},\ \bibinfo {pages} {461}
  (\bibinfo {year} {1976})}\BibitemShut {NoStop}%
\bibitem [{\citenamefont {Bhalerao}\ \emph {et~al.}(2015)\citenamefont
  {Bhalerao}, \citenamefont {Ollitrault}, \citenamefont {Pal},\ and\
  \citenamefont {Teaney}}]{Bhalerao:2014mua}%
  \BibitemOpen
  \bibfield  {author} {\bibinfo {author} {\bibfnamefont {R.~S.}\ \bibnamefont
  {Bhalerao}}, \bibinfo {author} {\bibfnamefont {J.-Y.}\ \bibnamefont
  {Ollitrault}}, \bibinfo {author} {\bibfnamefont {S.}~\bibnamefont {Pal}}, \
  and\ \bibinfo {author} {\bibfnamefont {D.}~\bibnamefont {Teaney}},\ }\href
  {\doibase 10.1103/PhysRevLett.114.152301} {\bibfield  {journal} {\bibinfo
  {journal} {Phys. Rev. Lett.}\ }\textbf {\bibinfo {volume} {114}},\ \bibinfo
  {pages} {152301} (\bibinfo {year} {2015})},\ \Eprint
  {http://arxiv.org/abs/1410.7739} {arXiv:1410.7739 [nucl-th]} \BibitemShut
  {NoStop}%
\bibitem [{\citenamefont {He}\ \emph {et~al.}(2016{\natexlab{a}})\citenamefont
  {He}, \citenamefont {Qian},\ and\ \citenamefont {Huo}}]{He:2016qjs}%
  \BibitemOpen
  \bibfield  {author} {\bibinfo {author} {\bibfnamefont {R.}~\bibnamefont
  {He}}, \bibinfo {author} {\bibfnamefont {J.}~\bibnamefont {Qian}}, \ and\
  \bibinfo {author} {\bibfnamefont {L.}~\bibnamefont {Huo}},\ }\href {\doibase
  10.1103/PhysRevC.93.044918} {\bibfield  {journal} {\bibinfo  {journal} {Phys.
  Rev.}\ }\textbf {\bibinfo {volume} {C93}},\ \bibinfo {pages} {044918}
  (\bibinfo {year} {2016}{\natexlab{a}})}\BibitemShut {NoStop}%
\bibitem [{\citenamefont {He}\ \emph {et~al.}(2016{\natexlab{b}})\citenamefont
  {He}, \citenamefont {Qian},\ and\ \citenamefont {Huo}}]{He:2016qwg}%
  \BibitemOpen
  \bibfield  {author} {\bibinfo {author} {\bibfnamefont {R.}~\bibnamefont
  {He}}, \bibinfo {author} {\bibfnamefont {J.}~\bibnamefont {Qian}}, \ and\
  \bibinfo {author} {\bibfnamefont {L.}~\bibnamefont {Huo}},\ }\href {\doibase
  10.1103/PhysRevC.94.034902} {\bibfield  {journal} {\bibinfo  {journal} {Phys.
  Rev.}\ }\textbf {\bibinfo {volume} {C94}},\ \bibinfo {pages} {034902}
  (\bibinfo {year} {2016}{\natexlab{b}})}\BibitemShut {NoStop}%
\bibitem [{\citenamefont {Niemi}\ \emph {et~al.}(2013)\citenamefont {Niemi},
  \citenamefont {Denicol}, \citenamefont {Holopainen},\ and\ \citenamefont
  {Huovinen}}]{Niemi:2012aj}%
  \BibitemOpen
  \bibfield  {author} {\bibinfo {author} {\bibfnamefont {H.}~\bibnamefont
  {Niemi}}, \bibinfo {author} {\bibfnamefont {G.}~\bibnamefont {Denicol}},
  \bibinfo {author} {\bibfnamefont {H.}~\bibnamefont {Holopainen}}, \ and\
  \bibinfo {author} {\bibfnamefont {P.}~\bibnamefont {Huovinen}},\ }\href
  {\doibase 10.1103/PhysRevC.87.054901} {\bibfield  {journal} {\bibinfo
  {journal} {Phys. Rev.}\ }\textbf {\bibinfo {volume} {C87}},\ \bibinfo {pages}
  {054901} (\bibinfo {year} {2013})},\ \Eprint {http://arxiv.org/abs/1212.1008}
  {arXiv:1212.1008 [nucl-th]} \BibitemShut {NoStop}%
\bibitem [{\citenamefont {Bzdak}\ \emph {et~al.}(2014)\citenamefont {Bzdak},
  \citenamefont {Bo\.zek},\ and\ \citenamefont {McLerran}}]{Bzdak:2013rya}%
  \BibitemOpen
  \bibfield  {author} {\bibinfo {author} {\bibfnamefont {A.}~\bibnamefont
  {Bzdak}}, \bibinfo {author} {\bibfnamefont {P.}~\bibnamefont {Bo\.zek}}, \
  and\ \bibinfo {author} {\bibfnamefont {L.}~\bibnamefont {McLerran}},\ }\href
  {\doibase 10.1016/j.nuclphysa.2014.03.007} {\bibfield  {journal} {\bibinfo
  {journal} {Nucl.Phys.}\ }\textbf {\bibinfo {volume} {A927}},\ \bibinfo
  {pages} {15} (\bibinfo {year} {2014})},\ \Eprint
  {http://arxiv.org/abs/1311.7325} {arXiv:1311.7325 [hep-ph]} \BibitemShut
  {NoStop}%
\bibitem [{\citenamefont {Bo{\.z}ek}\ \emph {et~al.}(2014)\citenamefont
  {Bo{\.z}ek}, \citenamefont {Broniowski}, \citenamefont {Arriola},\ and\
  \citenamefont {Rybczy\'nski}}]{Bozek:2014cva}%
  \BibitemOpen
  \bibfield  {author} {\bibinfo {author} {\bibfnamefont {P.}~\bibnamefont
  {Bo{\.z}ek}}, \bibinfo {author} {\bibfnamefont {W.}~\bibnamefont
  {Broniowski}}, \bibinfo {author} {\bibfnamefont {E.~R.}\ \bibnamefont
  {Arriola}}, \ and\ \bibinfo {author} {\bibfnamefont {M.}~\bibnamefont
  {Rybczy\'nski}},\ }\href {\doibase 10.1103/PhysRevC.90.064902} {\bibfield
  {journal} {\bibinfo  {journal} {Phys.Rev.}\ }\textbf {\bibinfo {volume}
  {C90}},\ \bibinfo {pages} {064902} (\bibinfo {year} {2014})},\ \Eprint
  {http://arxiv.org/abs/1410.7434} {arXiv:1410.7434 [nucl-th]} \BibitemShut
  {NoStop}%
\bibitem [{\citenamefont {Fu}(2015)}]{Fu:2015wba}%
  \BibitemOpen
  \bibfield  {author} {\bibinfo {author} {\bibfnamefont {J.}~\bibnamefont
  {Fu}},\ }\href {\doibase 10.1103/PhysRevC.92.024904} {\bibfield  {journal}
  {\bibinfo  {journal} {Phys. Rev.}\ }\textbf {\bibinfo {volume} {C92}},\
  \bibinfo {pages} {024904} (\bibinfo {year} {2015})}\BibitemShut {NoStop}%
\bibitem{Bialas:2007eg} 
  A.~Bialas and A.~Bzdak,
  Phys.\ Rev.\ C {\bf 77}, 034908 (2008)
  doi:10.1103/PhysRevC.77.034908
  [arXiv:0707.3720 [hep-ph]].
\end{thebibliography}

%

\end{document}